\documentclass[letterpaper]{article} 
\usepackage{aaai25}  
\usepackage{times}  
\usepackage{helvet}  
\usepackage{courier}  
\usepackage[hyphens]{url}  
\usepackage{graphicx} 
\usepackage{natbib}  
\usepackage{caption} 
\usepackage{algorithm}
\usepackage{algorithmic}

\usepackage{newfloat}
\usepackage{listings}

\usepackage{multirow}
\usepackage{graphicx}

\DeclareCaptionStyle{ruled}{labelfont=normalfont,labelsep=colon,strut=off} 
\floatstyle{ruled}
\newfloat{listing}{tb}{lst}{}
\floatname{listing}{Listing}
\nocopyright

\title{Exploring Socially Assistive Peer Mediation Robots\\ for Teaching Conflict Resolution to Elementary School Students}
\author{
    Kaleen Shrestha,
    Harish Dukkipati,
    Avni Hulyalkar,
    Kyla Penamante,
    Ankita Samanta,
    Maja Matarić
}
\affiliations{
    Department of Computer Science\\ Viterbi School of Engineering\\University of Southern California
}

\begin{document}

\maketitle

\begin{abstract}
In {\it peer mediation}--an approach to conflict resolution used in many K-12 schools in the United States--students help other students to resolve conflicts. For schools without peer mediation programs, socially assistive robots (SARs) may be able to provide an accessible option to practice peer mediation. We investigate how elementary school students react to a peer mediator role-play activity through an exploratory study with SARs. We conducted a small single-session between-subjects study with 12 participants.  The study had two conditions, one with two robots acting as disputants, and the other without the robots and just the tablet. We found that a majority of students had positive feedback on the activity, with many students saying the peer mediation practice helped them feel better about themselves. Some said that the activity taught them how to help friends during conflict, indicating that the use of SARs for peer mediation practice is promising. We observed that participants had varying reading levels that impacted their ability to read and dictate the turns in the role-play script, an important consideration for future study design. Additionally, we found that some participants were more expressive while reading the script and throughout the activity. Although we did not find statistical differences in pre-/post-session self-perception and quiz performance between the robot and tablet conditions, we found strong correlations ($p<0.05$) between certain trait-related measures and learning-related measures in the robot condition, which can inform future study design for SARs for this and related contexts.
\end{abstract}

\section{Introduction}
Interpersonal conflict is an inevitable part of every person’s life; knowing how to properly address and resolve conflict is a key life skill. Early practice of that skill is crucial to developing productive habits \cite{lane1992peer}. {\it Peer mediation} is a model of conflict resolution used in many schools across the United States that lets students help other students resolve conflicts, and is an empowering form of conflict resolution for both the mediators and the disputants \cite{johnson1996conflict, lane1992peer, maxwell1989mediation, angaran1999elementary}. Socially assistive robots (SARs), robots that help people through social rather than physical support \cite{deng2019embodiment}, are uniquely equipped to augment children’s social and emotional education \cite{shen2018stop}. For schools without peer mediation programs, SARs may be able to provide an accessible option to practice peer mediation. Past work in human-robot interaction (HRI) has explored robots helping to mediate conflict \cite{shen2018stop, jung2015moderate, stoll2018}; our work introduces a novel approach in which SARs were used as mock disputants to support practicing conflict resolution skills. Given the importance of self-esteem and self-perception in forming social relationships \cite{hosogi2012importance, humphries1999improving}, as well as the prevalence of role-play in peer mediator training \cite{block2012resolving, humphries1999improving}, we investigated how elementary school children responded to a role-play activity in which they served as peer mediators with or without SAR disputants. We conducted a small single-session between-subjects study with 12 participants divided into two conditions, one involving two SAR disputants, and the other involving no robots.\\
\indent We present preliminary qualitative and quantitative findings from an exploratory study. We found that majority of students had positive feedback on the activity, with many stating that the peer mediation practice helped them feel better about themselves. Some said that the activity taught them how to help friends during conflict, indicating that the use of SARs for peer mediation practice is promising. We observed that participants' varying reading levels affected their ability to read the script in the role-play activity, likely confounding some of the findings. While we did not find statistical differences in pre-/post-session change in self-perception and quiz performance between the robot and tablet conditions, we found strong correlations ($p<0.05$) between certain trait-related measures and learning-related measures in the robot condition. 

\section{Related Work}
\subsection{Multiple Robots in Human-Robot Interaction Design}
Our two-robot interaction design was inspired by research in multi-robot human interaction in HRI. \citet{leite2015emotional} explored using two SARs to narrate a story to a group of 1st and 2nd grade students and found promise in the use of multiple robots for social skill development. A study by \citet{vazquez2014spatial} showed that children were more engaged and attentive to the spoken elements of an interaction with a robot when a “side-kick” or second robot was present. In our study, in the robot condition, a student was presented with a conflict scenario between two SARs. As a peer mediator, the student was in the position to guide the SARs through the conflict, instead of being guided by the SARs. To the best of our knowledge, this is the first use of SARs for role-play for teaching students peer mediation.

\subsection{SARs for Child Social-Emotional Education}
Recent research has emphasized how SARs can act as a positive tool in K-12 education \cite{papadopoulos2020systematic}. \citet{blackburn2021use} argued that SARs are ``human enough" so children feel comfortable interacting with them while the fact that they are not human creates a judgement-free space for learning \cite{blackburn2021use, abu2024robot}. SARs can further enhance learning by increasing engagement, motivation, and curiosity \cite{papadopoulos2020systematic}. Our lab’s past work has demonstrated SAR-supported learning of both cognitive and social skills \cite{clabaugh2019long}. Furthermore, there is work in HRI demonstrating positive learning outcomes with robots as peers rather than tutors for children \cite{diyas2016evaluating, pareto2022children}, which is a motivating factor for SARs acting as disputants in this study.

\subsection{Computational Tools for Conflict Resolution}
In a Wizard-of-Oz study by \citet{shen2018stop}, a Keepon robot was set up alongside pre-school aged children playing in a group setting to resolve arguments that arose. The robot, controlled by a human, led children toward constructively resolving conflict by disrupting the argument flow with an attention-grabbing noise distraction \cite{shen2018stop}. \citet{jung2015moderate} conducted a Wizard-of-Oz study with university students where a robot intervened during conflict in a team exercises. They found that the robot intervention increased the participants' awareness of the conflict rather than the groups' tendency to suppress the conflict.  In contrast, our work uses the SARs as mock disputants rather than mediators.

\citet{shaikh2024rehearsal} use an LLM to automatically generate a conflict resolution training partner in a role-play activity. Their system guides the user through conflict scenarios and provides options for how to respond and feedback on user responses. The authors found that practice with Rehearsal doubled the use of cooperative strategies, emphasizing the potential of using LLMs as a tool in conflict resolution practice. The authors used an LLM to automatically generate a conflict scenario, while we used a pre-planned script to ensure student safety.  Furthermore, the use of SARs allows us to explore the role of nonverbal behaviors by both SARs and participants in the context of conflict resolution practice.

\section{Methods}
We designed an exploratory user study to explore several research questions.

\subsection{Research Question}
\textbf{RQ1}: How do elementary-school aged participants respond to peer mediation with SARs compared to a tablet? \\
\noindent \textbf{RQ2}: Are there differences in self-perception or learning outcomes between participants interacting with a SAR versus the tablet?\\
\noindent \textbf{RQ3}: Are any participant behavioral patterns or traits associated with participant self-perception and/or learning about peer mediation?
\subsection{Peer Mediation Role-Play Script}
The study session involved scripted role-play between the elementary school student participant who played a peer mediator and two hypothetical disputing students named Nova and Dali\footnote{We choose non-binary names with Hispanic roots to avoid gender bias and better relate to the majority Latino study participant population}. The disputants were embodied in SARs in one condition, and were virtually represented via the tablet interface in the other condition. The script we used was based on a pre-existing script by the Conflict Resolution Center of St. Louis \cite{script}, which we adapted to make it age-appropriate for our study.

\subsection{Peer Mediation Role-Play System}
We chose the updated Blossom robot for this study for its small size, usability, and friendly, zoomorphic design. Blossom is a 3D-printed robot based on an open-source platform originally developed by \citet{suguitan2019blossom}. \citet{shi2024build} further refined the design for better accessibility and affordability; this is the version we used for the two SARs in the robot condition. The robots were controlled by Raspberry Pi computers and had crocheted exteriors with button features; one was blue and the other gray, to facilitate differentiation of the two disputing characters. 

\begin{figure}
    \centering
    \includegraphics[width=\linewidth]{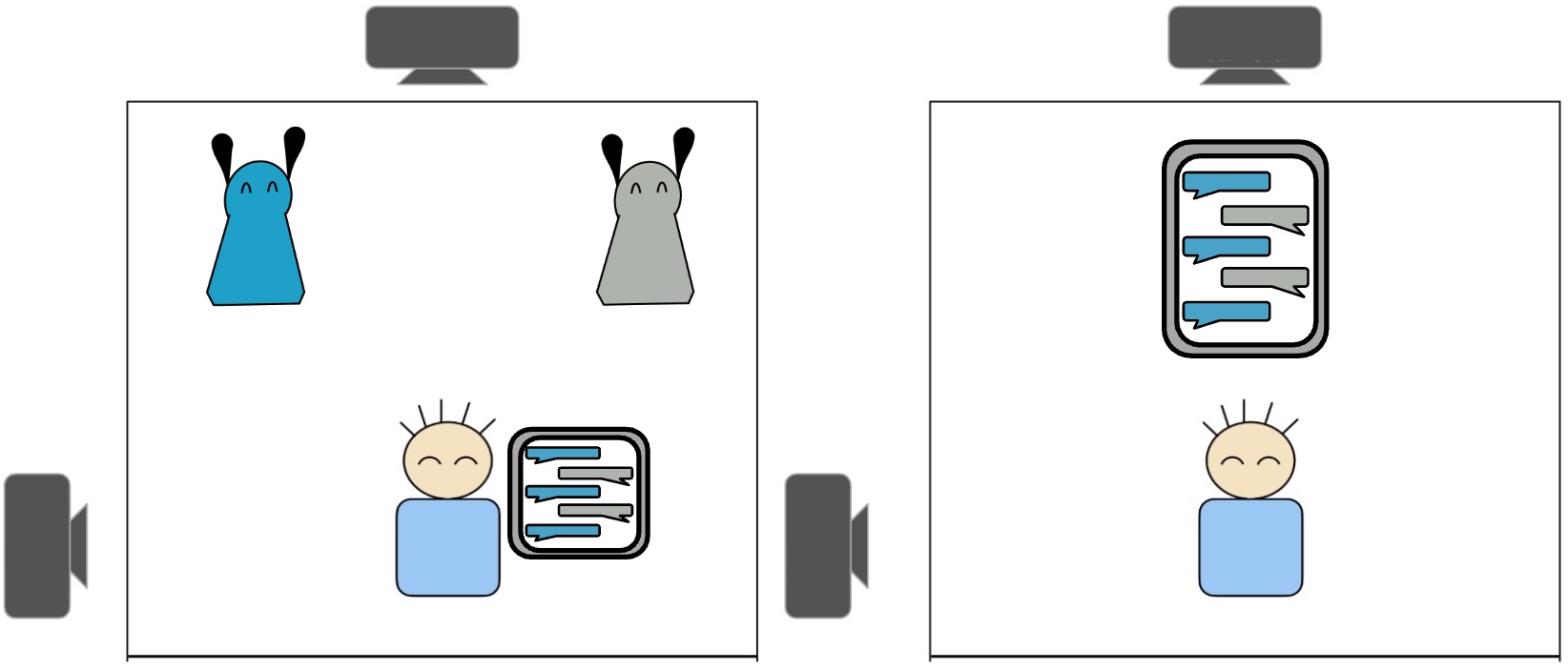}
    \caption{Study setup for robot condition (left) and tablet condition (right).}
    \label{fig:system-design-1}
\end{figure}

This system included a user interface (UI) displayed on a small Gechic portable 7.5 in. x 11 in. tablet-sized monitor (heretofore referred to as ``the tablet") that contained on-screen instructions for the participant, as well as a keyboard and mouse connected to the tablet for the participant to interact with the UI. The setup also included a tabletop tripod with a Logitech webcam across from the participant to capture  video data of the participant's face and audio data of the participant's speech, a tripod with a Logitech webcam setup to the left of the participant to capture video data of the full-body in profile. Figure \ref{fig:system-design-1} 
shows a schematic of the system setup for the two conditions of the study.

\subsection{Interaction Design}
During the role-play activity, each participant engaged in a verbal conversation with the disputants, serving as their peer mediator in a hypothetical conflict.  In the robot condition, the disputants were the two SARs, and in the tablet condition, they were represented by just the voice audio and on-screen images. The disputants had distinct voices that were clips pre-recorded with Narakeet, an online text-to-speech tool with child voices \cite{narakeet}. 

\begin{figure}
    \centering
    \includegraphics[width=\linewidth]{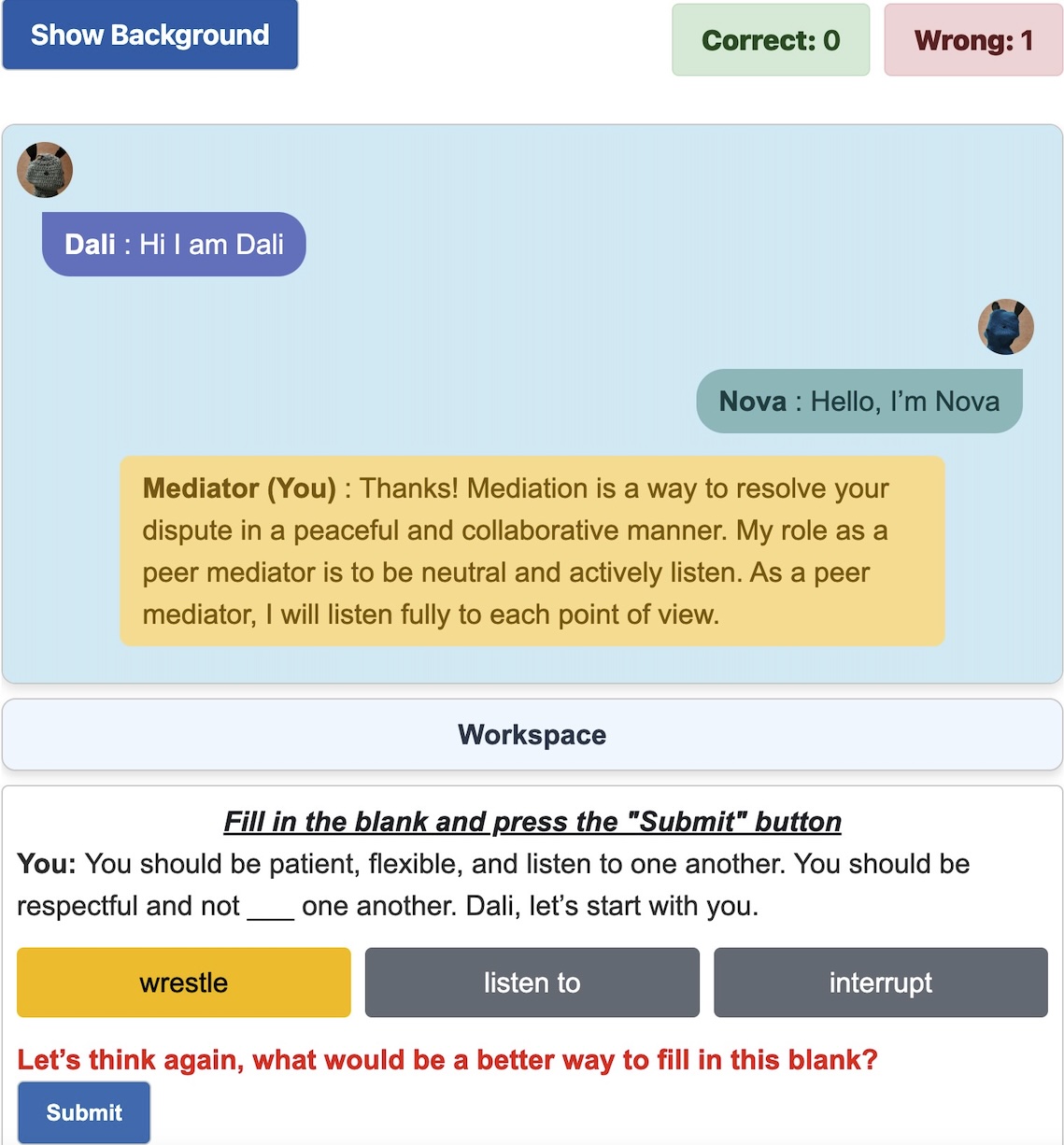}
    \caption{The UI displayed on the tablet monitor. The current turn is displayed (here, it is the participant's turn again, and they have just chosen the incorrect answer for the blank), and the current score.}
    \label{fig:ui-1}
\end{figure}

The participant spoke to the disputants in English and the disputants responded by speaking to the participant or to each other. In addition to audible speech, each turn of the conversation was displayed as text on the UI as well.  When it was the participant’s turn to speak, their script was also shown on the tablet UI. At some points in the session, portions of the participant's script were left blank and the participant was given three options to choose from to fill the blank, testing their mediation skills. If the participant did not choose the correct answer, the UI prompted them up to three times with an indirect hint (“Let’s think again, what would be the best way to fill in this blank?”) as was found to be effective \cite{lepper2013self}, up to three times. A counter displayed on the UI kept track of correct and incorrect answers, to motivate the participant and minimize guessing. Once the correct word or phrase was chosen, the UI prompted the participant to say the words aloud. At the end of the conversation, a summary was displayed on the UI giving the number of correct and incorrect answers. The conversational script was designed to eventually lead to a successful resolution of the conflict. 
Figure \ref{fig:ui-1} shows the UI.

Blossom robots were programmed to display movements that were appropriate and consistent with each robot's speech. We pilot tested a variety of movements that displayed emotion, such as strong head shakes for anger or slouching for sadness, with 3-5th grade students in a summer coding program held at our University. During pilot testing, we found that idle robot movements such as simulated breathing or swaying were distracting to students, so we excluded them; the robot's only moved while speaking.

\subsection{User Study}
\subsubsection{Study Design}
We conducted a preliminary single session between-subjects user study at a local elementary school with the following two conditions:\\
\indent \emph{Robot condition}: Participants were asked to complete the role-play activity with a tablet displaying the UI, mouse, keyboard, and two Blossom robots.\\
\indent \emph{Tablet condition}: Participants completed the role-play activity with a tablet displaying the UI, mouse, and keyboard. 
\subsubsection{Participants}
We recruited study participants by first recruiting an elementary school, by sending an informational flier. A small local elementary school expressed interest; the school had one class per grade. We encountered the usual challenges of working with K-12 students, including difficulties in contacting parents and obtaining consent, and constraints in scheduling the study due to the busy class schedule. Participant eligibility included being enrolled in 3-5th grade (ages 8-11), fluency in English, and ability to participate in person.  We recruited a small sample of 18 students, of whom two did not assent to participate. Of the 16 who assented, one participant had a technical failure during the session so their data were removed. Based on early insights, we simplified the script after the first day of the study so the two participants from the first day were also omitted from our analysis. Another participant, the only third-grade student, had difficulty reading the script and required significant support from the researcher, so their data were also omitted from the analysis. The results reported in this paper are based on the data from 12 participants who completed the study.  They identified as: 5 Male, 7 Female; 11 Hispanic/Latino, 1 White; ages ranged 9-11 ($M=9.9, SD=0.6$); grades ranged from 4th-5th. Students were randomly assigned conditions, with six students in the tablet condition and six students in the robot condition. Three participants in the robot condition reported having seen or used a robot before. One participant in the tablet condition reported being familiar with peer mediation prior to the study.

\subsubsection{Procedure}
The study was approved by the University Institutional Review Board (IRB \#UP-24-00495). We worked with the school to distribute parental consent forms to 3-5th grade classrooms. The researcher then reviewed an assent form with the student participants whose parents consented to the study. The participants assented to the study and completed a pre-session survey. The researcher then set up the system, and showed the participants how to use the system and complete the role-play activity for their condition. After the activity concluded, the participants completed a post-session survey and an open-ended feedback and demographics survey. The entire study duration took about 30 minutes ($M=34.5, SD=4.5$).

\subsection{Measures}
In the pre-session surveys, participants answered questions from the behavioral conduct subscale of the Self-Perception Profile for Children (SPP-C)
\cite{harter2012self}. They then completed the How I Respond to Conflict Worksheet (RCW) \cite{irex_conflict_response}. The researcher then gave a short lesson on conflict resolution and the participants completed a five question quiz on peer mediation based on the lesson. 

In the post-study surveys, participants repeated the SPP-C questionnaire and a quiz on peer mediation. They then completed the Pictorial Personality Traits Questionnaire for Children (PPTQ-C) \cite{mackiewicz2016pictorial}, which calculated personality scores based on the Big-5 personality framework \cite{costa1992four}. Finally, they answered questions about demographics, prior experience with peer mediation, prior experience with robots, and open-ended feedback about the activity.
\subsubsection{SPP-C: Self-Perception}
Introduced by \citet{harter2012self}, the behavioral conduct subscale is a six-question instrument that measures the degree to which one currently likes the way one behaves, does the right thing, acts the way one is supposed to act, and avoids getting into trouble. These concepts are related to self-regulation, which leads to better conflict resolution and is one of the main goals in peer mediation \cite{lane1992peer}. The scale for each item in this instrument ranges between 1 (low behavioral conduct self-perception) and 4 (high behavioral conduct self-perception).
\subsubsection{PPTQ-C: Personality}
The PPTQ-C measures five personality traits: extraversion, neuroticism, openness, conscientiousness, and agreeableness \cite{mackiewicz2016pictorial} based on the Big-Five framework \cite{costa1992five}, and is designed for children aged 7-13. The questionnaire includes 15 items, each with two pictures displaying opposite traits and prompts the participant to choose which picture they most closely relate to. Each item is rated on a 5-point scale ranging from “definitely yes” near the left side picture to “definitely yes” near the right side picture, with 1 being a low score for the particular personality trait tested by that item, and 5 being a high score. The score for each personality trait is calculated by averaging the relevant items and ranges between 1 (lowest) and 5 (highest). 
\subsubsection{RCW: Response to Conflict}
The RCW measures how participants typically react when faced with conflict. The worksheet consists of 14 responses to conflict that we categorize as cooperative (e.g., “Apologize”), avoidant (e.g., “Change the subject”), or competitive (e.g.,  “Call names”). Participants rate how often they exhibit each response on a 3-point scale ranging from 1 (Often) to 3 (Never). Finally, participants circle the three responses they normally show first in a conflict. We counted the number of cooperative, avoidant, and competitive responses from the three circled responses and used that score for further analysis.

\subsection{Distribution of Measures}
We present the distribution of personality, SPP-C scores (prior to the peer mediation activity), response to conflict preference, and total time taken for activity, for the sample populations in the robot and tablet conditions.
\subsubsection{Personality}
Table \ref{tab:personality-distribution} shows the average scores for each of the Big-Five personality dimensions. We also conducted independent samples t-tests and Mann-Whitney U tests between the two conditions for each dimension and found no significant differences between the two conditions for each personality dimension. The study participants were, on average, on the higher side of the agreeable, conscientiousness, and openness to experience dimension scales.

\begin{table*}[]
\centering
\begin{tabular}{llll}
\textbf{Big-Five Personality Trait} & \textbf{Robot}           & \textbf{Tablet}          & \textbf{t-test or Mann-Whitney U }               \\ \hline
Extraversion               & $M=3.8, SD=0.7$ & $M=3.0, SD=0.9$ & $U = 26.0, p=0.21$ \\
Neuroticism                & $M=2.6, SD=0.6$ & $M=2.7, SD=0.8$ & $t(10) = -0.1, p=0.91$  \\
Openness to Experience     & $M=3.9, SD=0.9$ & $M=3.8, SD=0.9$ & $t(10) = 0.1, p=0.92$ \\
Agreeableness              & $M=4.2, SD=0.9$ & $M=3.6, SD=0.8$ & $t(10) = 1.1, p=0.30$ \\
Conscientiousness          & $M=4.6, SD=0.5$ & $M=3.9, SD=0.6$ & $t(10) = 1.9, p=0.09$
\end{tabular}%
\caption{Distribution of personality scores for the five Big-Five personality traits for robot and tablet condition. As seen from the t-test results in the last columns, there was no significant difference between the two conditions for all traits.}
\label{tab:personality-distribution}
\end{table*}

\subsubsection{Top Three Responses to Conflict}
Table \ref{tab:rcw-categories} shows the average number of top three responses from the RCW that fall under the cooperative, avoidant, and competitive response categories. From the independent samples t-test and Mann-Whitney U test, we saw no statistically significant difference between the two conditions for each response category. We also saw that, on average, the majority of the responses chosen in the top three were in the cooperative category. 

\begin{table*}[]
\centering
\begin{tabular}{llll}
\textbf{Response to conflict type} & \textbf{Robot}  & \textbf{Tablet} & \textbf{t-test or Mann-Whitney U}           \\ \hline
Cooperative                        & $M=2.0, SD=1.2$ & $M=1.2, SD=1.1$ & $t(10) = -1.2, p=0.26$    \\
Avoidant                           & $M=0.5, SD=0.8$ & $M=1.0, SD=0.6$ & $U = 10.5, p=0.23$ \\
Competitive                        & $M=0.5, SD=0.8$ & $M=0.8, SD=0.7$ & $U=13.0, p=0.43$       
\end{tabular}%
\caption{Distribution of number of top three responses for each conflict response category for robot and tablet condition. As seen from the t-test and Mann-Whitney U test results in the last columns, there was no significant difference between the two conditions for all categories.}
\label{tab:rcw-categories}
\end{table*}

\subsubsection{Self-Perception Prior to Activity}
We found that there was not a significant difference in SPP-C scores for the robot condition ($M=3.1, SD=0.4$) and the tablet condition ($M=2.4, SD=0.6$): $t(10) = -2.1, p=0.06$. On average, our participants has SPP-C scores for behavioral conduct in the middle of the scale (around 2-3), indicating neither very high nor low self-perception of behavioral conduct.

\subsubsection{Time Taken for Activity}
We found that there was no significant difference in time taken (in seconds) for the full peer mediation role-play activity robot ($M=604.8, SD=162.6$) and tablet condition ($M=496.7, SD=135.0$): $t(10) = -1.14, p=0.28$. It took around 8-10 minutes to complete the activity.

\subsection{Analysis}
We implemented quantitative statistical tests using Pingouin version 0.5.5, a Python package \cite{Vallat2018pingouin}. To investigate the effect of embodiment on self-perception and learning of peer mediation, we calculated independent samples t-tests for change in SPP-C scores and number of questions correct on the peer mediation quiz for the two conditions. Specifically, we calculated independent sample t-tests over the differences between the pre- and post-session SPP-C scores (post-session SPP-C score - pre-session SPP-C score) over the robot and tablet conditions. Similarly, we calculated independent samples t-tests over the differences between the pre- and post-session peer mediation quiz number of questions correct (post number of quiz questions correct - pre number of quiz questions correct) over the two conditions. We used the Shapiro-Wilk test for normality and Levene test for equality in variances.

Next, we assessed the correlation between three trait-related measures and two measures related to learning efficacy. We assessed the Pearson correlation coefficient for the independent variable (trait-related measure) against the dependent variable (learning-related measure). We used the Shapiro-Wilk test to test for normality of the variable distributions. For variables that did not pass Shaprio-Wilk, we assessed correlation using Spearman's rank correlation coefficient.  
The two measures related to learning are:
\begin{enumerate}
    \item \textbf{Duration (in seconds)}: As discussed in Section 4.1.2, the pre/post-session peer mediation quiz scores did not demonstrate learning about peer mediation. This motivated us to analyze the time it took participants to do the reading and answer portions of the script. The reading level of each participant is one of the variables contributing to this measure; however, this measure can indicate how well a participant understands the activity and how careful the participant is in answering the questions in the script. For our analysis of each condition, we selected the turns with the highest variance in duration as an indication of difficulty..
    \item \textbf{Number of Attempts}: In each turn with the multiple choice question, participants were allowed up to three attempts for each such turn. We computed how many attempts were used to get to the correct answer as a measure of the participants' understanding of the peer mediation process and comprehension of the conversation.  
\end{enumerate}

We also collected qualitative observations from the feedback we received, and from the two video streams--facial and full-body profile--to inform future quantitative analysis.

\section{Results}
\subsection{Qualitative Results}
We found that when asked whether the activity made the participants feel better about themselves, eleven participants responded that the activity helped them feel better about themselves, with three responding that it helped a lot, and one responding that it helped a little bit. When asked if they learned to help friends, one participant wrote "Yes, because I see that some of my classmates argue and don't know how to handle it, so this could help." Another participant wrote "Yes, because I can either make new friends or help them be friends again." 

When asked to write what they thought of the peer mediation activity, all responded with positive comments that mentioned the activity being fun and helpful. The responses can be found in Table \ref{tab:mediation-thoughts}.

\begin{table*}[]
\centering
\begin{tabular}{ll}
\textbf{Condition} & \textbf{Thoughts About Peer Mediation Activity}                                   \\ \hline
tablet             & it was helpful because I learned I could help other students with this            \\
tablet             & it was a fun activity that i would do again                                       \\
tablet             & It was fun                                                                        \\
tablet             & It was fun, and I learned new things.                                             \\
tablet             & It helped me learn that I can help others instead of watching them fight.         \\
tablet             & It seems fun and helpful to others                                                \\
robot              & Helped me feel better.                                                            \\
robot              & I think that this is really fun to do and you can learn more about helping others \\
robot              & It's very fun \& helpful in case someone {[}illegible word{]}                     \\
robot              & I liked the peer mediation and the role-play                                      \\
robot              & I like it                                                                         \\
robot              & I was good                                                                       
\end{tabular}%
\caption{Responses to the open-ended question about participant's thoughts about the peer mediation activity.}
\label{tab:mediation-thoughts}
\end{table*}

We found that some participants were outwardly highly engaged, curious, and expressive, while others were more reserved, uninterested, or bored.  We are working on quantifying this difference in behavior by analyzing facial and postural features using OpenFace \cite{amos2016openface} and MediaPipe \cite{lugaresi2019mediapipe}.

\subsection{Quantitative Results}
\subsubsection{Embodiment and Self-Perception}\label{sec:embodiment-perception}
We found that there was not a significant difference in deltas of SPP-C scores between the robot condition ($M=-0.5, SD=0.5$) and the tablet condition $(M=0.11, SD=0.5)$: $t(10)= -2.2, p=0.053$. The p-value is close to significance of $0.05$, and so there is a slight possibility that there is a difference between the two conditions. The distribution of SPP-C delta scores is shown in Table \ref{tab:sppc_deltas}.  In the robot condition, there was a slight average decrease in SPP-C score after the session, with upwards of one point of decrease in SPP-C score. In the tablet condition, there is an average small increase in SPP-C score, with upwards of almost one point increase in SPP-C score. 

\begin{table}[]
\centering
\begin{tabular}{cc}
\multicolumn{2}{c}{\textbf{SPP-C delta (post - pre)}} \\ \hline
Tablet                     & Robot                    \\ \hline
-0.7                       & -1.2                     \\
-0.2                       & -0.8                     \\
0.2                        & -0.7                     \\
0.2                        & -0.5                     \\
0.3                        & -0.3                     \\
0.8                        & 0.3                     
\end{tabular}%
\caption{Change in self-perception after the session for both conditions. We see more slight decreases in self-perception after the peer mediation role-play activity.}
\label{tab:sppc_deltas}
\end{table}

\subsubsection{Embodiment and Peer Mediation Quiz Performance}\label{sec:embodiment-quiz}
We found that there was not a significant difference in deltas of number of correct answers in the quiz between the robot condition ($M=0.2, SD=0.7$) and the tablet condition ($M=0.2, SD=0.7$): $t(10)= 0.0, p=1.0$. For both conditions, there was a small average increase in the number of quiz questions (0.2 points increase on average) correct after the activity. In fact, for both conditions, we saw three participants with no improvement, two participants with one additional question correct, and one participant with one additional question incorrect after the peer mediation activity.

\subsubsection{Traits vs. Self-Perception}\label{sec:trait-perception} 
We report correlations between the three trait-related measures and change in self-perception.\\\\
\emph{Personality}
For each condition, we calculated the correlation between personality dimension scores and the SPP-C delta (post-pre SPP-C score). We did not find any significant correlations, although we found a near significant Pearson correlation coefficient for conscientiousness in the tablet condition: $r(6) = 0.79, p = 0.06$.\\

\noindent\emph{Response to Conflict Preference}
For each condition, we calculated the correlation between the conflict response preference categories and the SPP-C delta. We did not find any significant correlations in either condition.\\

\noindent\emph{Prior Self-Perception}
For each condition, we calculated the correlation between SPP-C scores prior to the activity and the SPP-C delta. We did not find any significant correlations in either condition.

\subsubsection{Traits vs. Learning Efficacy}\label{sec:trait-learning}
We report correlations between the three trait-related measures and time taken and number of attempts for turns in the activity.\\\\
\emph{Personality}
For each condition, we calculated correlation between personality dimensions scores and the duration of turns. We found significant strong Pearson correlation coefficients for extroversion, neuroticism, and conscientiousness in the robot condition for turn 3 of the script: extroversion ($r(6)=0.84$, $p=0.04$), neuroticism ($r(6) = -0.86$, $p=0.04$) and conscientiousness  ($r(6)=0.84$, $p=0.04$). For the tablet condition, we did not find any significant correlations. 

We then calculated correlation between personality dimensions scores and the number of attempts for turns with the highest variance of number of attempts. We found a significant strong negative Spearman's rank correlation coefficient for agreeableness in the robot condition for turn 5: $r(6)=-0.90$, $p=0.01$. For turn 18 in the robot condition, we found a significant strong positive Spearman's rank correlation coefficient for extroversion: $r(6)=0.88$, $p=0.02$. For the tablet condition, we did not find any significant correlations. \\

\noindent \emph{Response to conflict}
For each condition, we calculated the correlation between conflict response categories and the duration of turns with the highest variance in duration. For turn 3 in the robot condition, we found a significant strong negative Spearman correlation for avoidant response preference: $r(6)=-0.85$, $p=0.03$. We did not find significant correlations in the tablet condition.

We then calculated the correlation between conflict response categories and the number of attempts for turns with the highest variance of number of attempts. We found that for turn 5 in the robot condition, a significant strong negative Pearson correlation coefficient for cooperative conflict response preference: $r(6)=-0.87, p=0.03$. We did not find significant correlations in the tablet condition.\\

\noindent \emph{Prior Self-Perception}
For each condition, we calculated the correlation between SPP-C scores prior to the activity and the duration of turns with the highest variance. We did not find significant correlations in either condition.

For the correlations between SPP-C scores prior to the activity and the number of attempts, we did not find any significant correlations in either condition.

\section{Discussion and Future Work}
\subsection{RQ1: Response to Peer Mediation Activity}
While many participants thought the activity was enjoyable, reading skill was a challenge in our study design of the peer mediation role-play. In the future, we will look to limit the reading barrier in order to effectively teach children about conflict resolution in a personalized way. We plan to conduct co-design sessions and pilot studies with the intended participant population to diminish unexpected confounds.

Additionally, we plan to investigate affect and eye gaze from the study video and audio data to explore whether trait-related measures are related to expressivity and engagement of a participant since we found differences in expressiveness and attention during the activity. Insights from this exploration will help us to design more engaging and effective peer mediation scenarios and activities.

\subsection{RQ2: Embodiment and Self-Perception and Learning}
Due to the small sample size, we do not have enough evidence that the robot condition has significantly different pre-/post-session self-perception scores compared to the tablet condition. For both conditions, that there is only a small change (-0.5 for robot and 0.11 for tablet) in self-perception scores, indicating that self-perception did not change much after the single, short peer mediation session. 

Due to small sample size, we also do not have evidence that the robot condition has significantly different numbers of correct answers to the quiz pre-/post-session compared to the tablet condition. There is only a small change in quiz performance (0.2) for both conditions. This result could be caused by the short duration of the activity, the confounds within the activity, or by the quality of the quiz.

\subsection{RQ3: Participant Traits vs. Self-Perception and Learning Efficacy}
\subsubsection{Traits vs. Self-Perception} 
For personality, response to conflict preference, and self-perception measures prior to the session, we do not see significant correlations with change in self-perception for either condition. Changes in self-perception likely take longer than the duration of a short study session. Additionally our study sample was too small ($N=6$ per condition) to find any significant correlation coefficients. We did see a near-significant positive Pearson correlation coefficient of 0.79 between conscientiousness and change in self-perception in the tablet condition, suggesting a possible connection between carefulness and the peer mediation activity's ability to affect self-perception.  

As personality is known to play a key factor in creating effective interactions \cite{celiktutan2018computational, riggio1986impression}, we aim to investigate its relationship  with self-perception in a larger-scale study. We were also interested in seeing if there was any connection between preferences in conflict response and changes in self-perception (i.e., if a participant has a higher preference for competitive responses, would self-perception change after doing peer mediation that promotes cooperative responses), especially since the peer mediation role-play activity promoted cooperative responses to conflict. However, we did not see significant correlations with change in self-perception in our small single-session sample. Lastly, we were interested in the relationship between self-perception scores prior to the session and the change in self-perception, to investigate if, for example, lower initial self-perception leads to a greater change in self-perception after the session. We found no significant results for this variable.

\subsubsection{Traits vs. Learning Efficacy}
\begin{table*}[]
\centering
\begin{tabular}{lllll}
\multicolumn{1}{c}{\textbf{Turn \#}} & \multicolumn{1}{c}{\textbf{Prompt}}                                                                                                                                                                 & \multicolumn{1}{c}{\textbf{Option 1}}                                                              & \multicolumn{1}{c}{\textbf{Option 2}}                                  & \multicolumn{1}{c}{\textbf{Option 3}}                                        \\ \hline
3                                    & \begin{tabular}[c]{@{}l@{}}You should be patient, flexible, \\ and listen to each other. You should be \\respectful and not \_\_\_ each other. \\ Dali, let’s start with you.\end{tabular}               & interrupt                                                                                          & listen to                                                              & wrestle                                                                      \\\hline
5                                    & \begin{tabular}[c]{@{}l@{}}Ok, so you noticed Nova is playing with \\Andrea,who has always been mean to you. \\ After learning about this, you started \_\_\_\_. \\Does this sound right?\end{tabular} & \begin{tabular}[c]{@{}l@{}}making comments\\towards Nova about \\ how you were feeling\end{tabular} & \begin{tabular}[c]{@{}l@{}}ignoring Nova \\ during recess\end{tabular} & \begin{tabular}[c]{@{}l@{}}spreading false\\rumors about Nova\end{tabular} \\\hline
18                                   & \begin{tabular}[c]{@{}l@{}}Also, both of you want to \_\_\_, but continue \\to be respectful to each other. \\ What do you two think about this?\end{tabular}                                         &  \begin{tabular}[c]{@{}l@{}}move on and no longer\\be friends  \end{tabular}                                                                 & be friends again                                                       & play soccer together                                                        
\end{tabular}%
\caption{Participant turns from the peer mediation role-play script. Participants were asked to fill in the blank from the three presented options.}
\label{tab:turn-prompts}
\end{table*}
We discuss the results from the analysis between the three trait-related measures and time and number of attempts taken for turns in the activity. The script for the turns can be found in Table \ref{tab:turn-prompts}.\\\\
\emph{Personality} In the robot condition, we saw indications of a correlation between personality and the time taken during turns in the script. In particular, we see correlation between certain personality traits for turn 3 in the robot condition. Turn 3 is a statement of the rules for peer mediation, where the participant fills in the correct answer. In the context of the script, it comes after a turn that is also the participant's turn, rather than after a disputant turn. The large positive correlation coefficient for extraversion of $0.84$ for this turn indicates that more extraverted participants took longer reading and filling out that turn. This suggests that the activity was perhaps more challenging for more extraverted participants.  This is interesting because extraversion, the trait related to being social, does not seem to be directly related to comprehension. It could be that extroverted participants wanted to hear the disputant's response, but needed to adjust to being it being their turn again. The large negative correlation coefficient of $-0.86$ for neuroticism indicates more neurotic participants took less time reading and filling out that turn, which was surprising given that neuroticism is related to anxiety, worry, and other negative emotions and has been found to be negatively related to academic motivation \cite{apostolov2022role}. Lastly, we saw that conscientiousness had a high positive correlation to duration with a correlation coefficient of $0.84$, which is intuitive, since conscientiousness is related to being careful and meticulous. We did not see any relationship between personality and duration in the tablet condition, suggesting that personality interacts more strongly with SARs in the context of time taken for certain turns. 

We also saw indications of a correlation between personality and the number of attempts taken for a particular turn. Specifically, in the robot condition, we saw that for turn 5, agreeableness had a large negative correlation coefficient ($r(6)=-0.90$), indicating that more agreeable participants took less attempts on turn 5. Turn 5 was a recall question that tested comprehension of the conversation so far, and so agreeableness being relevant is surprising. It could be that agreeable participants want to cooperate with others, and thus try harder to answer the question correctly. In turn 18 in the robot condition, we saw that extraversion had a strong positive correlation coefficient ($r(6)=0.88$), indicating that more extraverted participants took more attempts for that turn. Turn 18 was again a recall question, and the correct answer was somewhat counterintuitive, with the conclusion that the best solution was for the disputants to move on and no longer be friends. The incorrect options were for the disputants to be friends again. More extroverted participants, characterized by being more sociable, warm, and enthusiastic, may be expected to choose a answer that aims for the disputants to be friends again, before arriving at the correct answer, perhaps resulting in the increase in attempts.  

We did not find any relationship between personality and number of attempts in the tablet condition, suggesting that personality may interact more strongly with embodied peers.  Since personality is known to play a role in learning styles and learning efficacy \cite{khatibi2016learning, ahmed2019impact}, our early findings present interesting directions for future studies with SARs. \\
\noindent\emph{Response to Conflict} In the robot condition, we saw indications of a correlation between conflict response categories and the time taken for reading/filling out turns in the script. In particular for turn 3 in the robot condition, we found that avoidant response preference had a large negative correlation ($r(6)=-0.85$). This means that participants that preferred avoidant responses more took less time completing turn 3, which is a statement of the peer mediation rules. We expected a large negative correlation for cooperative response preferences, since peer mediation is about cooperative responses, and so it is surprising avoidant response had the large negative correlation. It is also important to note that not many chose avoidant responses, and so we cannot be sure of this pattern even if we got significant correlation.  

We also saw indications of a relationship between conflict response categories and the number of attempts taken for a particular turn. Specifically, in the robot condition, we saw that for turn 5, we found a strong negative correlation for cooperative response preference ($r(6) = -0.87$). This means participants who prefer cooperative responses took less attempts on turn 5, which was a recall question about understanding a disputant's story. This pattern may be explained by the fact that peer mediation is an exercise in promoting cooperation. However, turn 5 was a recall question rather than a peer mediation question, and so it is surprising that conflict response preference had an association with comprehension. There may be a confounding variable that is related to comprehension and conflict response preference which is interesting future work.\\
\noindent \emph{Prior Self-Perception}
We saw no indication of a correlation between the participants' prior SPP-C scores and the time taken for reading/filling out turns in the script. This suggests that the time taken to read/fill out the turns in the peer mediation activity does not depend on the participant’s initial self-perception of behavioral conduct. Similarly, we saw no indication of a linear association between the participants' prior SPP-C scores and the number of attempts taken during certain turns in the script. This suggests that initial self-perception of behavioral conduct does not affect the comprehension of the peer mediation activity.



\section{Limitations}
In conducting the study, we found that reading skill may have been a significant confounding variable in the efficacy of the mock peer medication activity; the participants often read aloud slowly. Also, they may have choosen answers they thought were expected rather than what was true for them. For example, they may have chosen “understand each other” because “yell or threaten” seemed like an unacceptable choice.

Additionally, sometimes the only available times were during recess, likely impacting participant motivation.

Analyzing changes in self-perception is better suited for long-term studies. Our single session study was not able to capture such changes. Future work on peer mediation activities with SARs should ideally be implemented and evaluated over weeks of practice. 

\section{Acknowledgments}
We are grateful to Principal Sima Saravia-Perez of St. Odilia School for her support, and the teachers and students for their time and participation in this study. We also thank the summer high school student researchers from the Viterbi SHINE program for helping program the Blossom robot.

This material is based upon work supported in part by the NSF CISE Graduate Fellowship CSGrad4US under Grant No. 2313998. This work is also supported in part by the Viterbi CURVE Fellowship.
\bibliography{aaai25}
\appendix

\end{document}